\documentclass[aps,reprint, amsmath,amssymb]{revtex4-1}
\usepackage{amsmath}
\usepackage{amssymb}
\usepackage{graphicx}
\usepackage{hyperref}
\usepackage{bm}
\usepackage{url}
\usepackage[usenames,dvipsnames]{xcolor}
\usepackage{tabularx}
\usepackage{array}
\usepackage{colortbl}

\begin{document}
\bibliographystyle{plain}
\title{Effect of VSR invariant Chern-Simon Lagrangian on photon polarization }

\author{Alekha C. Nayak}
\email{acnayak@iitk.ac.in}
\author{Ravindra K. Verma}
\email{ravindkv@iitk.ac.in}
\author{Pankaj Jain}
\email{pkjain@iitk.ac.in}

\affiliation{Department of Physics, Indian institute of Technology, Kanpur\\ Kanpur, India- 208016}

\begin{abstract}
We propose a generalization of the Chern-Simon (CS) Lagrangian which is
invariant under the SIM(2) transformations but not under the full Lorentz
group. 
We study the effect of such a term on radiation propagating 
over cosmological distances. We find that the dominant effect of this
term is to produce 
 circular polarization as radiation propagates through space.
We use the circular polarization data from distant radio sources in
order to impose a limit on this term.
\label{sec:abstract}
\end{abstract}

\maketitle
\section{Introduction}

\label{sec:intro}
Symmetry is the guiding principle of modern physics. The Standard Model 
(SM), which is based on Lorentz invariance,
 provides a successful description of nature and has passed 
all experimental tests. However, it is believed to be the
 low energy limit of an ultimate theory at the Planck scale energy. At 
this scale, unification of
 quantum field theory with gravity leads to possible 
violation of Lorentz symmetry \cite{Collins:2004bp}. However it is difficult to probe Lorentz violation 
at LHC because the signal is supressed by ratio of electroweak scale ($M_W$) to 
Planck mass ($M_{Pl}$), i.e. $M_W/M_{Pl} 
\approx 10^{-17}$ \cite{Colladay:1998fq}. 
In a supersymmetric (SUSY) theory the signal may be suppressed more strongly
as $(M_{SUSY}/M_{Pl})^2$, where $M_{SUSY}$ is the scale of SUSY breaking
\cite{GrootNibbelink:2004za,Jain:2005as}.

 Several terrestial \cite{PhysRev.42.400}, astrometric \cite{bay,Müller199571} and astrophysical tests have been conducted on Lorentz violation which impose
stringent limits on its violation. Different theoretical models have been proposed on possible depatures from Lorentz invariance \citep{Kostelecky:2009zp,Bluhm:2004ep,Berger:2001rm,Carroll:2001ws,Colladay:2001wk,Colladay:1996iz,Kostelecky:2010hs,Coleman:1998ti,
Collins:2004bp,GrootNibbelink:2004za,Jain:2005as}. Colladay and 
Kostelecky \citep{Colladay:1998fq} consider the Standard Model Extention (SME) in which Lorentz symmetry is spontaneously violated. 
Carroll et al. \citep{Carroll:1989vb} consider a CS term in 3+1 dimension, which is gauge invariant but breaks Lorentz invariance. The authors introduce 
 an external four vector which breaks Lorentz invariance. 
This term is local and rotates the plane of polarization of photon due to different velocities of left- and right-circularly polarized photon.\\
                            
  We consider SIM(2) invariant CS term which shows partial gauge invariance
 but breaks Lorentz invariance. SIM(2) is the proper subgroup of Lorentz group as developed by Cohen and Glashow \citep{Cohen:2006ky} and termed 
Very Special Relativity (VSR). Adjoining one of the discrete symmetry such as P, T, CP or CT with Lorentz subgroup enlarges it to the full Lorentz group. 
In this paper, we introduce nonlocal operator  $\frac{n_\alpha}{n\cdot\partial} $ in the CS term in 3+1 dimension. This operator violates Lorentz invariance while respecting SIM(2) invariance. As we shall see, 
the modified nonlocal CS term splits 
the photon into two different polarization states which travel with different velocities. This implies violation of parity and Lorentz invariance in the theory. We find that the dominant effect of this term is to generate
circular polarization as the electromagnetic wave travels through space. 
Furthermore the effect is dominant at low frequencies.
Using the circular polarization data of radio sources from the MOJAVE 
(Monitoring of Jets in active galactic nuclei (AGN) with Very Long Baseline Array (VLBA) Experiments) program \citep{Homan:2005nu}, we impose a 
limit on the Lorentz violating parameter. 
  
   This paper is organised as follows: in Sec. \eqref{sec:theory} we present the VSR invariant CS Lagrangian and derive the resulting photon dispersion relation. We then obtain the formulas for the Stokes parameters in this model. In Sec. \eqref{sec:result} we extract the Lorentz invariance violating parameter 
by the standard $\chi^2$ minimization procedure. 
Finally, we conclude in Sec. \eqref{sec:conclusion}.

\section{Theory}
\label{sec:theory}
The well established Maxwell's theory of electrodynamics is 
based on gauge and Lorentz invariance. The Lagrangian density for 
massless photons is given by
\begin{equation}
  \mathcal{L} = -\frac{1}{4}F_{\alpha\beta}F^{\alpha\beta}
\label{eq:em1}
\end{equation}
where $F_{\alpha\beta}=\partial_\alpha A_\beta-\partial_\beta A_\alpha$ is electromagnetic tensor. Eq. \eqref{eq:em1} is invariant under the gauge transformation $A_\alpha\rightarrow A_\alpha+\partial_\alpha \varphi $. 
A photon mass term breaks 
the gauge invariance of the Lagrangian and experimental data imposes 
a stringent limit on this term.
We propose a SIM(2) invariant nonlocal CS term which can be written as,  
\begin{equation}
  \mathcal{L}_{cs} = \frac{\Gamma}{2} \frac{n_\alpha}{n\cdot\partial}A_\beta \tilde{F}^{\alpha\beta}
\label{eq:cs}
\end{equation}
Here
 $\tilde{F}^{\alpha\beta}$ is dual electromagnetic tensor, $\Gamma$ is a 
parameter of dimension mass square and $n^\alpha=(1,0,0,1)$. This is manifestly not Lorentz invariant \citep{Cohen:2006ir} but is invariant 
under SIM(2) transformations. The corresponding generators are
$T_1=K_x+J_y, T_2=K_y-J_x$, rotations ($J_z$) and boosts ($K_z$) about z-axis. 
Here $\bf{J}$ and $\bf{K}$ are the generators of rotations and boosts respectively.
 Under a boost along z-axis ($K_z$), the preferred 
vector $n$ transforms 
as $n^\alpha\rightarrow e^\phi n^\alpha$. However $\frac{n_\alpha}{n\cdot\partial} $ is homogeneous in $n$, and hence is SIM(2) invariant.

Under the gauge transformation $A_\alpha\rightarrow A_\alpha+\partial_\alpha \varphi $, the variation of SIM(2) modified CS term is
\begin{eqnarray}
&& \Delta\mathcal{L^\prime}_{cs}= \frac{\Gamma}{2} \left(\frac{n_\alpha}{n\cdot\partial}\right)\partial_\beta \varphi\tilde{F}^{\alpha\beta} \nonumber \\
&& =\frac{\Gamma}{2}  \partial_\beta\left(\frac{n_\alpha}{n\cdot\partial}\varphi\tilde{F}^{\alpha\beta} \right)  
\label{eq:gauge}
\end{eqnarray}
where we have used the fact that $n^\alpha$ is a constant vector and 
$\partial_\beta \tilde{F}^{\alpha\beta}=0$.
The remaining term is a surface term and would normally give null contribution
to the action. However in the present case, this term also involves the 
inverse of $n\cdot\partial$. We find that this term vanishes in all cases
except if the derivative $\partial_\alpha$ is taken in the direction of
$n^\alpha$. In this case the derivative operation cancels with the inverse
operator $1/n\cdot\partial$ and we obtain a finite contribution. 
In order to eliminate this violation of gauge invariance we impose the
constraint,
\begin{equation}
n\cdot A = 0
\end{equation} 
This constraint is not invariant under Lorentz transformations but 
is invariant under SIM(2) transformations \cite{2012PhRvD..85j5009V}. 
Hence we can consistently impose it within our framework. We also point
out that the vector field $A_\mu$ does not form an
irreducible representation of
SIM(2) and hence it is not necessary for us to work with the full vector
field. In fact the vector potential can be split into
four independent one dimensional SIM(2) fields \cite{2012PhRvD..85j5009V}.

The complete Lagrangian density in presence of conserved current $J^\alpha$ is
\begin{equation}
\mathcal{L}_{T} = -\frac{1}{4}F_{\alpha\beta}F^{\alpha\beta}- J^\alpha A_\alpha+ \frac{\Gamma}{2} \frac{n_\alpha}{n\cdot\partial}A_\beta \tilde{F}^{\alpha\beta}
\label{eq:lagra}
\end{equation}
The equation of motion for the Lagrangian density $\mathcal{L}_{T}$ is 
\begin{equation}
\partial_\beta F^{\beta\alpha} = J^\alpha - \Gamma \frac{n_\lambda}{n\cdot\partial}\tilde{F}^{\lambda\alpha}
\label{eq:eom}
\end{equation}
The modified Maxwell's equations from Eq. \eqref{eq:eom} are given by 
\begin{subequations}
\begin{eqnarray}
&& {\bf{\nabla}}\cdot {\bf{E}} = \rho + \Gamma\frac{1}{n\cdot\partial} {\bf{n}}\cdot{\bf{B}}\label{subeq:max1}\\
&& -\partial_t {\bf{E}}+ {\bf{\nabla}}\times{\bf{B}}= {\bf{J}}+ \Gamma\frac{1}{n\cdot\partial} ({\bf{B}}- {\bf{n}}\times{\bf{E}})\label{subeq:max4}
\end{eqnarray}
\label{eq:max}
\end{subequations}
Here $\bf{n}$ is a unit vector along the $z$ axis.
The two homogeneous Maxwell's equations are
\begin{subequations}
  \begin{eqnarray}
    &&\hspace{20pt} {\bf{\nabla}}\cdot{\bf{B}} = 0\label{subeq:max2}\\
&&\hspace{20pt} {\bf{\nabla}}\times{\bf{E}} = - \frac{\partial{\bf{B}}}{\partial t}\label{subeq:max3}
  \end{eqnarray}
  \label{eq:hmmax}
\end{subequations}

Using Eqs. \eqref{subeq:max3}  and \eqref{eq:max}, the source free wave equation takes the form
\begin{equation}
 {\bf\nabla(\bf\nabla\cdot{\bf{E}})}-{\bf\nabla^2\bf{E}}+\frac{\partial^2{\bf{B}}}{\partial t^2}=\Gamma\frac{\partial}{\partial t}\left(\frac{1}{n\cdot\partial} {\bf{B}}\right)-\Gamma\frac{\partial}{\partial t}\left(\frac{1}{n\cdot\partial} {\bf{n}\times \bf{E}}\right)
 \label{eq:dis}
\end{equation}
The operator $\frac{1}{n\cdot\partial}$ becomes, 
\begin{equation}
\frac{1}{n\cdot\partial}=\frac{1}{\partial_t+\partial_z}=\int dt_+
\label{eq:wv}
\end{equation}
where $t_+=\frac{t+z}{2}$.
Eq. \eqref{eq:dis} further simplifies to 
\begin{equation}
(\omega^2-k^2){\bf{E}}+({\bf{k}}\cdot{\bf{E}}){\bf{k}} = \frac{\Gamma}{\omega-k\cos\theta}({{\bf{k}}}\times{\bf{E}}-\omega{\bf{n}}\times{\bf{E}} )
\label{eq:eom2}
\end{equation}
In the present case, for the source free modified Maxwell's equation ${\bf{k}}\cdot{\bf{E}}\neq 0$, so the longitudinal component of the photon polarization
(proportional to $\Gamma$) is not zero. It acquires a small value
compared to the transverse components. Since the vector $\bf{n}$ is along the 
z-direction 
and the wave propagation vector, ${\bf{\hat{k}}}$, makes an angle $\theta$ with 
$\bf{n}$, the electric field can be expressed as, 
\begin{equation}
 {\bf{E}} = E_x{\bf{\hat{x}}}+E_{yz}{\bf{\hat{p}}} +E_k{\bf{\hat{k}}}
\end{equation}
where
\begin{subequations}
\begin{eqnarray}
&& {\bf{\hat{p}}}= -{\bf{\hat{y}}}\cos\theta +{\bf{\hat{z}}}\sin\theta\label{subeq:vec2}\\
&& {\bf{\hat{k}}} = {\bf{\hat{z}}}\cos\theta +{\bf{\hat{y}}}\sin\theta\label{subeq:vec3}\\
&& {\bf{\hat{n}}}={\bf{\hat{z}}}\label{subeq:vec4}
\end{eqnarray}
\label{eq:wave}
\end{subequations}
Comparing the x-, y- and z- component of Eq. \eqref{eq:eom2}, we get
\begin{subequations}
  \begin{eqnarray}
 &&\hspace{-20pt}(\omega^2-k^2)E_x =\nonumber\\ && \hspace{8pt}\frac{\Gamma}{(\omega-k\cos\theta)} \{ (k-\omega\cos\theta)E_{yz}+\omega\sin\theta E_k\} \\
&& \hspace{-20pt}\Gamma E_x = (\omega^2-k^2)\cos\theta E_{yz}-\omega^2\sin\theta E_k\\
&&\hspace{-23pt} \frac{\Gamma k\sin\theta}{\omega-k\cos\theta}E_x=-(\omega^2-k^2)\sin\theta E_{yz}-\omega^2\cos\theta E_k  
  \end{eqnarray}
  \label{eq:comp}
\end{subequations}
Using Eq. \eqref{eq:comp}, we get the following dispersion relation
\begin{equation}
\omega^2- k^2 =\pm i \Gamma 
\label{eq:dispers}
\end{equation}
Here the + and $-$ signs correspond to right- and left-handed polarized 
photons. 
Hence we find that 
the modified SIM(2) invariant CS term leads to different dispersion relations
for the right and left handed polarizations.

From Eq. \eqref{eq:comp}, the relation between the components of electric fields 
for the two solutions in Eq. \ref{eq:dispers} are given by
\begin{subequations}
\begin{eqnarray}
&& E_x \approx \mp i \frac{(\omega -k\cos\theta)}{(k-\omega\cos\theta)}E_{yz}\\
&& E_k \approx \frac{(\omega^2 -k^2)\sin\theta}
{\omega^2(1-\cos\theta)}E_{yz}
\end{eqnarray}
\end{subequations}
where we have kept only the leading order terms in $\Gamma$. 
Using Eq.  \eqref{eq:dispers}, we find that the two eigenmodes of
propagation are, 
\begin{equation}
|E_+\rangle \approx \left(
\begin{array}{cc}
\frac{(\omega-k\cos\theta)}{(k-\omega\cos\theta)}\\ 
i\\
\frac{i(\omega^2 -k^2)\sin\theta}{\omega^2(1-\cos\theta)}
\end{array}   
\right)
,
|E_-\rangle \approx \left(
\begin{array}{cc}
\frac{(\omega-k\cos\theta)}{(k-\omega\cos\theta)}\\
-i\\
\frac{-i(\omega^2 -k^2)\sin\theta}{\omega^2(1-\cos\theta)}
\end{array}   
\right)
\label{eq:eigenvectors}
\end{equation}
In the limit $\omega \approx k$ the corresponding eigenvalues are given by
\begin{equation}
k \approx \omega \mp \frac{i \Gamma}{2\omega} \equiv k_\pm
\label{eq:dispers2}
\end{equation}
 where we have used Eq. \eqref{eq:dispers}.  

In the limit, $\Gamma \rightarrow 0$, the two vectors in Eq. \ref{eq:eigenvectors} correspond to the left and right circular polarizations, i.e.
\begin{equation}
|E_+\rangle = \left(
\begin{array}{cc}
1\\ 
i\\
0
\end{array}   
\right)
,
|E_-\rangle= \left(
\begin{array}{cc}
1\\ 
-i\\
0
\end{array}   
\right)
\end{equation}
 Eq. \eqref{eq:dispers2} implies that two polarization modes travel with different velocity which is an indication of parity violation. This model leads to
a significant contribution to circular polarization at low frequencies, 
which we will discuss in the next section.

Let us label the axes along $\bf{\hat x}$, $\bf{\hat p}$ and $\bf{\hat k}$
by $x_1$, $x_2$ and $x_3$ respectively. Hence our wave is propagating
along the $x_3$ direction and the two transverse directions are taken
to be along $x_1$ and $x_2$. 
An electric state vector, at any given time, can be written as a linear combination of two state vectors given in Eq. \eqref{eq:eigenvectors}.i.e 
\begin{eqnarray}
|E(x_3,t)\rangle = \frac{E_+(0)}{\sqrt{2}} \left(
\begin{array}{cc}
\frac{(\omega-k_+\cos\theta)}{(k_+-\omega\cos\theta)}\\ 
i\\
\frac{i(\omega^2 -k_+^2)\sin\theta}{\omega^2(1-\cos\theta)}
\end{array}  
\right)e^{i(k_+x_3-wt)} \nonumber \\
&& \hspace{-170pt}+ \frac{E_-(0)}{\sqrt{2}} \left(
\begin{array}{cc}
\frac{(\omega-k_-\cos\theta)}{(k_--\omega\cos\theta)}\\ 
-i\\
\frac{-i(\omega^2 -k_-^2)\sin\theta}{\omega^2(1-\cos\theta)}
\end{array}  
\right)e^{i(k_-x_3-wt)}
\label{eq:linear}
\end{eqnarray}
Here $E_3(x_3,t)$ component of the state vector $|E(x_3,t)\rangle$ is very small and we are interested in determining the change in the photon polarization in the plane perpendicular to the photon propagation. 
Using Eq. \eqref{eq:linear}, we obtain, 
\begin{subequations}
\begin{eqnarray}
&&\hspace{-30pt} E_1(x_3,t)=\frac{1}{\sqrt{2}}(P_+E_+(0) e^{i k_+ x_3} \nonumber \\
&& \hspace{50pt}+ P_-E_-(0) e^{ik_- x_3})e^{-i\omega t}\label{subeq:Ex}\\
&& \hspace{-30pt} E_2(x_3,t)=\frac{1}{\sqrt{2}}(iE_+(0) e^{i k_+ x_3} \nonumber \\
&& \hspace{50pt} - iE_- (0)e^{ik_- x_3})e^{-i\omega t}\label{subeq:Ey}
\end{eqnarray}
\label{Field}
\end{subequations}
where
\begin{equation}
P_\pm = \frac{ \omega-k_\pm\cos\theta}{k_\pm -\omega\cos\theta}
\end{equation}
The electric vector rotates in the plane perpendicular to the direction of propagation, when the two photon polarization modes travel with different velocity. 
Hence, the polarization state of photon changes after propagation over
a large distance. 
It can be determined by calculating the Stokes parameter I,Q,U,V.

We assume that the wave is unpolarized at source and calculate 
its polarization after propagation through a distance $x_3$. The Jones matrix for unpolarized electromagnetic wave is  
\begin{equation}
J= \left(
\begin{array}{cc}
1 & 0\\ 
0 & 1
\end{array}   
\right)
\label{eq:J0unpol}
\end{equation} 
Using this as the  initial condition, we calculate the Stokes parameters 
after the photon has travelled a distance $x_3$. 
 These are given by,
\begin{subequations}
\begin{eqnarray}
I =&& 2 \cosh \left(\frac{\Gamma x_3}{\omega }\right)\\
Q =&& 0 \\
U= && \frac{2 \Gamma \cot ^2\left(\frac{\theta }{2}\right) \sinh ^2
\left(\frac{\Gamma x_3}{2 \omega }\right)}{\omega ^2}\\
V =&& 2 \sinh \left(\frac{\Gamma x_3}{\omega }\right)
\end{eqnarray}
\label{eq:stokes}
\end{subequations}
Keeping only the leading order in $\Gamma$, we obtain, 
\begin{equation}
I=2,Q=0,U=0,V=\frac{2\Gamma}{\omega}x_3
\label{eq:st2}
\end{equation}
This implies that an initially unpolarized wave acquires circular polarization
upon propagation. The polarization state does not depend upon the direction of propagation of photon with respect to the VSR preferred axis.
From Eq. \eqref{eq:st2}, we obtain,
\begin{equation}
  \xi\equiv \frac{V}{I}=\frac{\Gamma}{\omega}x_3 
  \label{eq:y}
\end{equation}

So far we have confined our analysis to a flat space-time. However
we need to compute the change in polarization in an expanding Universe since
we are interested in sources located at redshifts comparable to unity.
We consider a spatially flat Universe. The propagation of electromagnetic
wave follows the same equations as given above but with time replaced
by conformal time and the distance $x_3$ 
replaced by comoving distance \cite{1991PhRvD..43.3789C,2012PhRvD..86k5025T}. 
Besides this 
the overall energy density in the wave decreases due to expansion. However
this effect is not relevant for calculation of shift in polarization. 
Consider a source at a redshift $z$. Its comoving distance $x_3$ is given 
by \citep{Weinberg:2008zzc} 
\begin{equation}
  x_3=\frac{1}{a_{0}H_{0}}\int_{\frac{1}{1+z}}^1 \frac{\mathrm{d}x}{x^2\sqrt{\Omega_{\Lambda}+\Omega_{M}x^{-3}}}
\label{eq:redshift}
\end{equation}
where $H_{0}$ is Hubble constant, $a_{0}$ is scale parameter at present epoch, $\Omega_{\Lambda}$ is the ratio of vacuum energy density to critical density and $\Omega_{M}$ is the ratio of non-relativistic 
matter density to critical density. Hence Eq\eqref{eq:y} becomes
\begin{eqnarray}
   \xi\equiv\frac{V}{I}=\frac{\Gamma}{\omega a_{0}H_{0}}\int_{\frac{1}{1+z}}^1 \frac{\mathrm{d}x}{x^2\sqrt{\Omega_{\Lambda}+\Omega_{M}x^{-3}}} \nonumber \\
&& \hspace{-120pt} =\beta\int_{\frac{1}{1+z}}^1 \frac{\mathrm{d}x}{x^2\sqrt{\Omega_{\Lambda}+\Omega_{M}x^{-3}}}
\label{eq:redshift1}
\end{eqnarray}
where $\beta=\frac{\Gamma}{\omega a_{0}H_{0}}$ is dimensionless. 
We find that the circular polarization depends on the redshift of the
source.
Furthermore the polarization generated increases with decrease in frequency. 
Hence the
effect is dominant at low frequencies, such as, radio waves. In comparison
the dominant effect in the case of the local Lorentz violating term, studied in
Ref. \cite{Carroll:1989vb} is to produce a frequency independent rotation
of linear polarization. We point out that mixing of hypothetical pseudoscalars
of very low mass with photons in a background
magnetic field also generates circular polarization 
\cite{1983PhRvL..51.1415S,1986PhLB..175..359M,1988PhRvD..37.1237R,2002PhRvD..66h5007J,2005JCAP...06..002D}. In this case the
effect is found to increase with frequency and is limited by the
stringent constraints that 
have been imposed on the circular polarization which may be generated
at optical frequencies \cite{2012JCAP...07..041P}.

In more generality, the theoretical model can be expressed as, 
\begin{equation}
\xi=\beta\int_{\frac{1}{1+z}}^1 \frac{\mathrm{d}x}{x^2\sqrt{\Omega_{\Lambda}+\Omega_{M}x^{-3}}}+{\xi}_{0}
\label{eq:y2}
\end{equation}
where we have added a constant
 $\xi_0$ to the model
in order to allow for the possibility that the mean value of circular 
polarization over the sample is not zero. In the limit $\beta=0$, 
$\xi_0$ is the mean value of circular polarization over the entire sample.
For $\beta\ne 0$ we determine both the parameters, $\beta$ and $\xi_0$, by
making a $\chi^2$ fit to data.

\section{Results}
\label{sec:result}
In this section, we determine the best fit parameters by using
 the $\chi^2$ minimisation procedure for the circularly polarized light 
emitted from radio jets associated with AGN. 
                Here we use the circular polarization data from the MOJAVE program, which contains 133 bright, mostly compact radio-loud AGN in the northern sky. With VLBA facilities, circular polarization of the AGN jet sample at 15
Ghz with flux density 1.5 Jy has been observed. We only consider sources
with redshift $z$ 
greater than 0.25 for which local effects are absent. 
We find that after this cut only 102 sources remain in the data set.

\begin{figure}[h]
\begin{center}
\includegraphics[scale=0.30, angle=270]{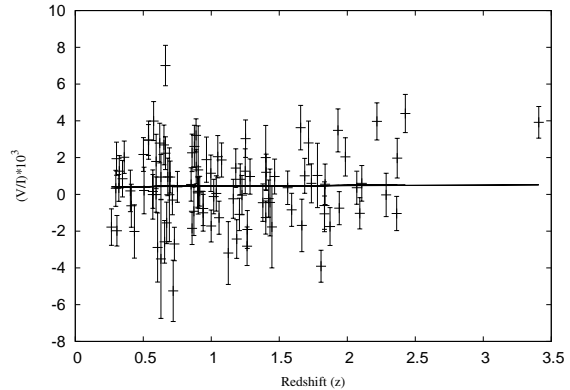}
\caption{The fit to the circular polarization data from distant radio
galaxies. }
\label{fig}
\end{center}
\end{figure}

We extract the Lorentz 
violating parameter ($\beta$) by minimizing $\chi^2$ for the model 
given in Eq. \ref{eq:y2}. The resulting $\chi^2$ is compared with the
null model 
 $\xi={\xi}_{0}$ in order to determine the significance of the fit.  
For the null model we find that the mean value $\xi_0=0.38$ with 
$\chi^2=347.6$. The Lorentz violating model, Eq. \ref{eq:y2}, yields
$\beta=0.067\pm 0.12$, $\xi_0=0.40$ with minimum value of $\chi^2=347.1$. Hence
we do not find a significant signal of Lorentz violation
in the data. 
The corresponding best fit is shown in Fig. \eqref{fig}.
Using the extracted value of $\beta$ we find that the one sigma limit on
the Lorentz violating parameter $\Gamma$ is $(13\times 10^{-29}\ \text{GeV})^2$. 



\section{Conclusion}
We have proposed a modified Chern-Simon term which breaks the full Lorentz 
invariance but is invariant under SIM(2) transformations. The term is non-local
and depends on a prefered vector, $n^\alpha$. The term also respects 
gauge invariance provided we impose the condition, $n\cdot A=0$, 
on the gauge potential, $A_\alpha$. This condition is invariant under
the SIM(2) transformation and can be imposed consistently in this theory.
The non-local CS term changes the dispersion relation of photon. 
Hence it changes the polarization of the electromagnetic waves travelling
over large distances. We find that at leading order,
an initially unpolarized picks up circular polarization upon propagation. 
We test the predicted signal by using the 
 circular polarization data from distant radio galaxies. We do not
find a significant signal of violation of Lorentz invariance and
impose a stringent limit on the Lorentz violation parameter. 
\label{sec:conclusion}

\bibliographystyle{apsrev}
\bibliography{Ref_vsr}

\end{document}